\documentclass[doublecol]{epl2} 

\DeclareRobustCommand\openone{\leavevmode\hbox{\small1\normalsize\kern-.33em1}}
\def\be{\begin{equation}}
\def\ee{\end{equation}}

\usepackage{graphicx}

\begin{document}
\title{Quantum phase transition between cluster and antiferromagnetic states}

\author{Wonmin Son \inst{1} \and Luigi Amico \inst{2} \and Rosario Fazio \inst{1,3}, Alioscia Hamma \inst{4} \and Saverio Pascazio \inst{5,6} \and Vlatko Vedral \inst{1,7,8}}
\shortauthor{W. Son \etal}

\institute{
\inst{1} Center for Quantum Technology, National University of Singapore, 117542 Singapore, Singapore\\
\inst{2} CNR-MATIS-IMM $\&$ Dipartimento di Fisica e  Astronomia Universit\`{a} di Catania, C/O ed. 10, viale A. Doria 6 95125 Catania, Italy\\
\inst{3} NEST, Scuola Normale Superiore \& CNR-INFM , Piazza
	dei Cavalieri 7, I-56126 Pisa, Italy\\
\inst{4} Perimeter Institute for Theoretical Physics, 31 Caroline St. N,Waterloo ON, N2L 2Y5, Canada\\
\inst{5} Dipartimento di Fisica and MECENAS, Universit\`a di Bari, I-70126 Bari, Italy\\
\inst{6} INFN, Sezione di Bari, I-70126 Bari, Italy\\
\inst{7} Department of Physics, National University of
	Singapore, 2 Science Drive 3, Singapore 117542\\
\inst{8} Department of Physics, University of Oxford, Clarendon Laboratory, Oxford, OX1 3PU, UK
}


\abstract{
We study a Hamiltonian system describing a three spin-$1/2$ cluster-like interaction competing with an Ising-like exchange. We show that the ground state in the cluster phase possesses symmetry protected topological order. A continuous quantum phase transition occurs as result of the competition between the cluster and Ising terms. At the critical point the Hamiltonian is self-dual. The geometric entanglement is also studied. Our findings in one dimension corroborate the analysis of  the two dimensional generalization of the system, indicating, at a mean field level, the presence of a direct  transition between an antiferromagnetic and a valence bond solid ground state.
}

\maketitle

\section{Introduction}
The Landau theory of phase transitions is at the heart of our understanding of critical phenomena~\cite{goldenfeldbook,sachdevbook}. By introducing the concept of order parameter Landau's theory relates phase transitions to symmetry breaking. The order parameter (e.g.,  magnetization in ferromagnets) is different from zero in the less symmetric (ferromagnetic) phase while it vanishes in the paramagnetic phase where symmetry is restored. However, not all quantum phases of matter can be classified according to their symmetries~\cite{Wenbook} and  there are quantum phase transitions that elude Landau's paradigm~\cite{deconfined,sachdev-review}. These phases, dubbed topological phases, cannot be characterized by a local order parameter. Topological phases play a prominent role in very diverse physical contexts, ranging  from the quantum Hall effect and high $T_C$ superconductivity in solid state physics to confinement problems in QCD and string theory~\cite{Wenbook}. The continuous (direct) transition between a Neel antiferromagnet and a valence bond solid belongs to such class of exotic phase transitions~\cite{deconfined}. The transition occurs because a new conserved quantity (a gauge symmetry) appears exclusively at the critical point, characterized by the emergence of certain topological defects in the zero temperature antiferromagnet which become energetically favorable  and  proliferate (i.e.\ they are
deconfined). At the critical point within the Landau-Ginzburg scheme, such phase transitions must be of the first order and they cannot be continuous.

Understanding exotic quantum phases and quantum phase transitions lying outside Landau's paradigm are among the most intriguing
and controversial topics in the modern theory of critical phenomena~\cite{deconfined-firstorder,sachdev-report,fradkin}. Exact results,
desirable for a thorough comprehension of such phenomena,
are extremely rare in the field.
In this Letter we discuss an exotic phase transition between symmetry protected topological order and an antiferromagnetic state occurring in an exactly solvable one dimensional model. Based on the one dimensional theory we produce evidence that in two dimensions our model displays a continuous quantum phase transition between an antiferromagnet and a valence bond solid. Our model can be experimentally realized in optical lattices and ion traps (see the concluding remarks) and is relevant in quantum information (in certain limits the ground state is a cluster state, central in one-way quantum computation~\cite{Briegel01}).

\section{The model}
The Hamiltonian system that we shall study emerges as a result of the cross fertilization  between quantum statistical mechanics and quantum computation.
It reads
\begin{equation}
	H(\lambda)=-\sum_{i=1}^N \sigma_{i-1}^x\sigma_{i}^z\sigma_{i+1}^x +\lambda \sum_{i=1}^{N}
	\sigma_{i}^y\sigma_{i+1}^y ,
\label{eq:hamiltonian}
\end{equation}
where $\sigma_i^\alpha$ is the Pauli matrix acting on the $i$-th site of a $1$-D lattice. (Similar models were considered in~\cite{Doherty09,Skrovseth09}.) This describes the interplay between a cluster and an antiferromagnetic Ising Hamiltonian.  The model can be defined also in higher dimensions and the nature of its phase transition is of great interest. First we focus on its properties in one-dimension where the model can be solved
exactly. For $\lambda = 0$ and open boundary conditions, the ground state (gs) manifold is given by
\begin{equation}
	\mathcal C_0 =\mbox{span} \left\{ \frac{1}{2^N}(\sigma^x_1 )^k(\sigma^x_L)^l \prod_{i} \hat{C}_i
	|0\rangle \; \; k,l=0,1\right\}
	\label{eq:C0}
\end{equation}
where $\hat{C}_i=\left (\openone
	-\sigma_i^x-\sigma_{{\cal N}(i)}^x-\sigma_i^x \sigma_{{\cal N}(i)}^x \right )$,
${\cal N}(i)$ is the oriented nearest neighbours of the lattice site $i$ and $\sigma^z|0\rangle = |0\rangle$. Such definition is valid for a lattice of arbitrary dimension. The gs manifold is thus four-fold degenerate. With periodic boundary conditions the gs is unique and is given by $k=l=0$.
 Cluster states are a special type of  multi-qubit graph states~\cite{Briegel01}  with a computational power
that is believed to be an important resource for measurement based quantum computation~\cite{Nielsen05}. In this context it is relevant to study
how the  computational properties of the cluster states are quenched under additional perturbation. This kind of questions have been recently studied
in a series of works~\cite{Doherty09,Skrovseth09} where it was shown that the cluster phase is unstable above a critical value of certain one- and
two-sites (Ising-like) perturbations. The computational power of cluster states can be viewed as a quantification of their entanglement~\cite{Damian07, Vedral08}.
Localizable entanglement, in particular, was calculated and it was shown that its range diverges~\cite{Skrovseth09},  a property shared with
spin half valence bond states~\cite{cirac-localizable,camposvenuti}.  Interestingly enough the two states are both characterized by a hidden order of topological nature. Very recently it is emerging that long-range entanglement is a inherent property of topologically ordered states \cite{Wen-protected}.
Inspired by the problem posed in~\cite{deconfined} we will prove that the model
(\ref{eq:hamiltonian}) undergoes to a  continuous transition between an Ising antiferromagnet and a phase characterized by  the specific topological order encoded in  cluster states. We will show
that entanglement captures the critical properties of this model, {beyond  the formulation  of the concept of order parameter.  The two dimensional generalization
of the model will be discussed at the end of this letter.

\section{Symmetry protected topological order in cluster states}
Topological order is defined as a degeneracy of the ground state depending on the topology of the system.
Moreover, such degeneracy must be robust under arbitrary local perturbations, in the sense that for a finite system of linear size $L$ the splitting of the degeneracy is exponentially small in $L$.
Recently, it has been understood that the notion of topological order described above cannot be applied to  $1D$ \cite{Wen-protected}. For one dimensional systems and therefore also for the $1D$  cluster state, it results that  the degeneracy is not robust under generic perturbations. Indeed, for the state (\ref{eq:C0})  the four basis states are connected by the local operators $\sigma^x_1, \sigma^x_L$ and distinguished by the local operators $\sigma^z_1 \sigma^x_2$ and $\sigma^x_{L-1} \sigma^z_L$. Nevertheless, a more tenuous kind of topological order can be possible, if symmetries are present. If not all operators or perturbations are allowed, but only those that respect some symmetry, the symmetry can protect the topological order. In the present case, the ground state manifold  $\mathcal C_0$ is also defined by the algebra of the global operators $(X_1, Z_1)$ and $(X_2,Z_2)$, that are defined as
\begin{eqnarray}
\nonumber
X_1 =\prod_{n=0}^{\frac{L-8}{6}} ( \sigma_{6n+1} ^y \sigma_{6n+2} ^z  \sigma_{6n+3}^z \sigma^y_{6n+4}  \sigma_{6n+5} ^x \sigma_{6n+6} ^x ) \sigma_{L-2}^y\sigma_{L-2}^z\sigma_{L}^z \\
\nonumber
Z_1=\prod_{n=0}^{\frac{L-8}{6}} ( \sigma_{6n+1} ^x \sigma_{6n+2} ^x  \sigma_{6n+3}^y \sigma^z_{6n+4}  \sigma_{6n+5} ^z \sigma_{6n+6} ^y ) \sigma_{L-2}^x\sigma_{L-2}^x\sigma_{L}^y \\
\nonumber
X_2=\prod_{n=0}^{\frac{L-8}{6}} ( \sigma_{6n+1} ^x \sigma_{6n+2} ^y  \sigma_{6n+3}^z \sigma^z_{6n+4}  \sigma_{6n+5} ^y \sigma_{6n+6} ^x ) \sigma_{L-2}^x\sigma_{L-2}^y\sigma_{L}^z \\
Z_2=\prod_{n=0}^{\frac{L-8}{6}} ( \sigma_{6n+1} ^y \sigma_{6n+2} ^x  \sigma_{6n+3}^x \sigma^y_{6n+4}  \sigma_{6n+5} ^z \sigma_{6n+6} ^z ) \sigma_{L-2}^y\sigma_{L-2}^x\sigma_{L}^x 
\end{eqnarray}
These operators obviously square to the identity: $X_1^2=X_2^2=Z_1^2=Z_2^2=I$ and, if $L = 3(2k+1)$, with $k$ integer, the following anticommutation relations hold: $\{X_1,Z_1\}=\{X_2,Z_2\}=0$ and commute otherwise, thus realizing two copies of the Pauli algebra which constitutes the logical operators in the ground space $\mathcal C_0$. The Hamiltonian with $\lambda =0$ has thus a $Z_2\times Z_2$ global symmetry {defined by $T_i = X_i+Z_i, i=1,2$ that protects the topological degeneracy in the ground space.  This means that every perturbation $V$ that commutes with the symmetries is not able to split the degeneracy unless there is a QPT. This indicates that  there is a whole phase in which the ground state degeneracy is protected. The ising term though, violates this symmetry, so as soon as we switch on the $\lambda$ coupling the degeneracy is lifted.} In the limit of large $\lambda$, there is only the usual $Z_2$ symmetry and the system is antiferromagnetic. Aspects related to the preservation of topological order  
will be further invetigated in a forthcoming article \cite{cya}.

The properties of  $\mathcal C_0$ reflect the emergence of low energy states localized at the ends of an
{\it open chain}~\cite{boundary,kitaev-leshouches}; such phenomenon can be detected by the
so called  string order parameter, originally employed to study the Haldane phase displayed by one dimensional integer-spin systems~\cite{haldane}.  Affleck-Kennedy-Lieb-Tasaki models provide a paradigmatic example in this context (see \cite{AKLT} for a recent reference), with symmetry protected topological ordered ground states \cite{Wen-protected,AKLT,AKLT-protected}. We observe that (\ref{eq:C0}) is symmetric for alternating spin flips, $\sum_{k}(-1)^k \langle C_0|\sigma^y_k |C_0\rangle=0$, and any local order is absent. The string order parameter ${\cal O}_z=(-)^{N-2}\langle \sigma_1^y\prod_{j=1}^{N-1} \sigma_j^z \sigma_N^y\rangle \neq 0$, reflects the symmetry breaking of the state by $\pi$ rotations about the $y$ and the $z$ axes. It can be shown that ${\cal O}_z=1$ for the
cluster state, i.e.\ the gs of (\ref{eq:hamiltonian}) at $\lambda=0$. The gs of the antiferromagnetic Ising Hamiltonian $|I_0\rangle$ corresponds to the $\lambda\rightarrow\infty$ limit, instead. The  symmetry of the system is spontaneously broken, so that the staggered  magnetization $M=\sum_{k}(-1)^k \langle I_0| \sigma^y_k|I_0\rangle $ is an extensive  quantity in the thermodynamic limit.

\section{Duality mapping}
Duality transformations provide a powerful tool to extract properties of the phase diagram beyond the perturbative regions~\cite{Savit80}. We now apply the duality transformation
\begin{equation}
\label{eq:dual}
	\mu_{i}^z =\sigma_i^x\sigma_{i+1}^x, ~~~
	\mu_{i}^x =\prod_{j=1}^{i}\sigma_{j}^z.
\end{equation}
to our Hamiltonian. Observe that Eq.\ (\ref{eq:dual}) is a unitary transformation.

We first consider the case $\lambda=0$ in Eq.\ (\ref{eq:hamiltonian}).
For open boundary condition, $\sigma_{0}^x=\sigma_{N+1}^x=1$, the  cluster
Hamiltonian $H(\lambda=0)$ is transformed into an Ising Hamiltonian with a boundary term:
\begin{equation}
\label{eq:Transformed}
	H(\lambda=0)\rightarrow H_I=\sum_{i=1}^{N-1}
	\mu_{i}^y\mu_{i+1}^y-\hat{B} ,
\end{equation}
where $\hat{B}= i
(\prod_{i=1}^{N}\mu_i^z) \mu_{1}^y$.
Being unitary, the dual mapping (\ref{eq:dual}) factors out the correlation terms into a boundary term.
The gs of (\ref{eq:Transformed}) is a superposition of states with anti-ferromagnetic order,
$|G\rangle=(|I_0\rangle+|\bar{I}_0\rangle)/\sqrt{2}$,
with $|I_0\rangle=|+_y,-_y,+_y,-_y
\cdots\rangle$, $|\bar{I}_0\rangle\equiv
\hat{B}|I_0\rangle$,
$\mu^y|\pm_y\rangle=\pm|\pm_y\rangle$, and
$H_I|G\rangle=-N|G\rangle$, where $-N$ is the non-degenerate lowest eigenvalue of the Hamiltonian.
$|G\rangle$ can be transformed into a GHZ state
$|\textrm{GHZ}\rangle=(|+\rangle^{\otimes N}+|-\rangle^{\otimes N})/\sqrt{2}$
by local unitary transformations. Therefore, after the dual
transformation, the cluster state is transformed into a GHZ state and
the unit of entanglement is degraded from $N/2$ to $1$ \cite{Vedral08}.
On the other hand, for periodic boundary conditions, Eq.\ (\ref{eq:dual}) yields 
\begin{eqnarray}
\label{eq:dual-hamiltonian2}
H(\lambda=0)\rightarrow H_{I,p} = \sum_{i=1}^{N-2}
\mu_{i}^y\mu_{i+1}^y-\hat{B}_p ,
\end{eqnarray}
where $\hat{B}_p=
(\prod_{i=1}^{N}\mu_i^z)( i \mu_{1}^y\mu_{N}^z-\mu_{N-1}^y\mu_{N}^y)$.
The gs is again a GHZ state $|G_p\rangle=(|I_0\rangle_{N-1}+|\bar{I}_0\rangle_{N-1})
\otimes |+_x\rangle/\sqrt{2}$ and the same considerations apply.

We now tackle the case $\lambda\neq 0$.  The dual model of (\ref{eq:hamiltonian}) is
\begin{eqnarray}
\label{eq:dual-hamiltonian}
H(\lambda)_{\rm{dual}}=\sum_{i=1}^{N-2} \mu_{i}^y\mu_{i+1}^y
-\lambda \sum_{i=1}^{N-1}
\mu_{i-1}^x\mu_{i}^z\mu_{i+1}^x-\hat{B},
\end{eqnarray}
where $\mu_{0}^x=1$ and $\hat{B}=
(\prod_{i=1}^{N}\mu_i^z)( i \mu_{1}^y\mu_{N}^z-\mu_{N-1}^y\mu_{N}^y+
i \mu_{N-1}^x\mu_N^y\mu_1^x)$. 
In the thermodynamic limit
$N \rightarrow \infty$, the duality relation reads
$H(\lambda)_{\rm{dual}}=\lambda H(\lambda^{-1})$,
since the boundary can be neglected \cite{Lieb61}.
The self-duality relation at $\lambda=1$ signals the critical point.

By resorting to the duality properties (\ref{eq:dual-hamiltonian}) 
the local order parameter of the antiferromagnet is mapped into the string order parameter of the cluster states (see also~\cite{Doherty09,Skrovseth09}). In particular we note  that the $\sum_i\langle \sigma^x_{i-1}\sigma^z_i \sigma^x_{i+1} \rangle/N $ cannot be taken as a valid order parameter because it enjoys the same symmetries of the Hamiltonian.

\section{Fermion representation and QPT}
The model defined here can be diagonalized by resorting to a
Jordan-Wigner transformation $c_l^{\dagger}=\prod_{m=1}^{l-1} \sigma_m^z
\sigma_l^{+}$, $c_l^\dagger c_l=\sigma_l^z+1/2$ (spinless fermions), yielding
\begin{eqnarray}
\label{eq:fermihamiltonian}
\nonumber
	H(\lambda)&=& \sum_{l=1}^{N} (c_{l-1}^\dagger-c_{l-1})(c_{l+1}^\dagger+c_{l+1}) \\
&+& \lambda \sum_{l=1}^{N} (c_{l}^\dagger+c_{l})(c_{l+1}^\dagger-c_{l+1}), 
 \end{eqnarray}
with boundary conditions $c_{N+l}=\mp c_{l}$. $N$ is assumed to be even.
We note that the three spin interaction of the cluster Hamiltonian is reflected in the \emph{next-nearest} neighbor
interaction in the fermionized Hamiltonian (\ref{eq:fermihamiltonian}).
The dispersion is obtained  by Fourier transformation $c_l =\sum_k e^{2\pi i k l/N} b_k/\sqrt{N}$
followed by a Bogoliubov transformation $b_k = u_k \gamma_k + v_k \gamma_{-k}^{\dagger}$, yielding:
\begin{eqnarray}
\nonumber
H(\lambda) &= & \sum_{k} \Lambda_k (\gamma_k^{\dagger}\gamma_k-1/2), \qquad \\ \Lambda_k 
&=& \sqrt{(\lambda^2 +1)-2\lambda \cos\left(6\pi k/ N\right)},
\end{eqnarray}
with $u_k=z_k^+$,
$v_k=-i \mbox{sign}(\delta_k)z_k^-$,
$z_k^\pm=\sqrt{(1\pm\epsilon_k/\Lambda_k)/2}$,
$\epsilon_k=[\cos(4\pi k/N)-\lambda\cos(2\pi k/N)]$, $\delta_k=[\sin(4\pi k/N)+\lambda\sin(2\pi k/N)]$ and
$-(N/2+1)/2<k<-1$, $N=4n+2$ [$-(N-1)/2<k<-1/2$, $N=4n$] for [anti]periodic boundary conditions.
The gs is identified as the vacuum of the Bogoliubov operators $\gamma_k|\Omega\rangle=0$, $\forall k$.
In terms of the $b_k$'s it can be written as the BCS ground state
\begin{eqnarray}
|\Omega\rangle =\prod_{k}
\left(u_k+v_k b^{\dagger}_k
b^{\dagger}_{-k}\right) |vac\rangle.
\end{eqnarray}
From simple inspection of the dispersion law of $\Lambda_k$, one infers that the system displays
a second order quantum phase transition at $\lambda=1$ with
critical indices $z=\nu=1$. Notice also that even though any two-spin correlations vanish in the cluster states \cite{Nielsen05, Bartlett06},
the fermionic pair correlations pattern is not trivial.

Entanglement is a key quantity in the study of many-body systems \cite{Amico08}. We now focus on how maximal entanglement in the cluster state \cite{Damian07} is demoted by nearest-neighbor interaction, as $\lambda$ increases in model (\ref{eq:hamiltonian}).
We  consider a geometric measure of entanglement \cite{Wei03} that provides a global characterization of the entanglement in quantum many-body systems
\begin{equation}
\label{eq:geomatic measure}
\varepsilon\left(|\psi\rangle\right)=
-\log_2\left[\max_{\theta} |\langle
S(\theta)|\psi\rangle|^2\right],
\end{equation}
where $|S(\theta)\rangle$ is the closest separable state. $N$-partite pure separable states are characterized by $2N$ real parameters,
\begin{equation}
\label{eq:sepst}
|S(\{\bar{\theta}_j\})\rangle =\prod_{j=1}^N \left(\cos \theta_j +e^{i\phi_j}\sin\theta_j \sigma_j^x\right)|\uparrow\rangle^{\otimes N},
\end{equation}
that can be restricted to four because of the dimerized structure of the eigenstate of the Hamiltonian.

\begin{figure}[ht]
\begin{center}
\includegraphics[width=8.5cm]{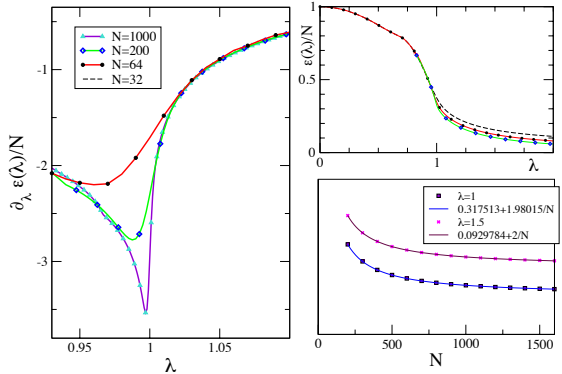}
\end{center}
\caption{Left panel: The first derivative of the geometric entanglement captures  the critical point at $\lambda=1$; $N$ ranges from $32$ (black-dashed line) to $1000$ (turquoise-triangle line). Right upper pannel:  Normalized geometric gs entanglement  $\epsilon/N$ as function of $\lambda$.  The decay of entanglement in the Ising phase is shown in the lower right panel: squares and crosses are the data for $\epsilon/N$  at the critical point and in the antiferromagnetic phase, respectively (an artificial offset has been applied for a convenient display).
}
\label{fig:ground}
\end{figure}

The geometric entanglement (\ref{eq:geomatic measure}) is plotted in Fig.\ \ref{fig:ground}.
In the right upper panel the normalized geometrical entanglement is plotted as a function of $\lambda$ for different values of $N$. For $\lambda=0$, the geometric entanglement is always
$N/2$. In the cluster phase ($\lambda < 1$) entanglement is very weakly dependent on the size of the system, while in the antiferromagnetic phase ($\lambda > 1$) it decreases with $N$.
Moreover, as shown in the right lower panel, although the normalized entanglement is very small for $\lambda>1$, it does not vanish.
Observe that the convexity of the geometric entanglement changes from positive to negative near the critical point. As $N$ is increased, the point of convexity change moves toward $\lambda=1$. This is scrutinized in the left panel, where one sees that the phase transition is detected by the derivative of the geometric entanglement, that diverges at $\lambda = 1$.

\section{$d$-dimensional systems}
The model defined by the Hamiltonian (\ref{eq:hamiltonian}) can be generalized to $d$-dimensions, the cluster term involving  the product of $2d+1$ spin operators (we consider for simplicity a hyper-cubic lattice).  We do not rely on exact results in this case; nevertheless we will show that the system has a critical value of the parameter $\lambda$ below which the antiferromagnetic order
vanishes.  This argument, together with the remarkable observation due to Cirac and Verstraete~\cite{VBS-cluster} that cluster states can be realized as
valence bond states, leads us to conclude  that our model displays  a {\it direct transition between an antiferromagnet and a valence bond state, for $d>1$}.  We comment that the treatment we are going to employ is mean-field in nature and although it is not expected to capture the correct critical behavior, it suffices to establish  the  very existence of the transition point.

The mean-field analysis \cite{negele-orland} proceeds after  a rotation of the Hamiltonian to map the antiferromagnetic to ferromagnetic Ising couplings. The approximated Hamiltonian reads:
\begin{equation}
\label{eq:Hmf}
H_{\rm{MF}}=-\sum_{i}\sigma^z_i \prod_{j\in{\cal N}(i)} \sigma^x_{j} - 2d \lambda \psi \sum_{i}\sigma^y_i \doteq H_C+V
\end{equation}
where $\psi\equiv\langle \sigma^y_i\rangle$ is the (real and spatially uniform) order parameter and $2d$ the coordination number of the lattice.
By assuming that the order parameter vanishes at the critical point, the self-consistency equation leads to the condition
\begin{equation}
\label{eq:sccond}
1=4 d \lambda_c \sum_m \int^{\infty}_0 d\tau e^{-(\epsilon_m-\epsilon_0)\tau} |\langle C|\sigma^y_i|C_{m}\rangle|^2 ,
\end{equation}
where $|C_m\rangle$ and $\epsilon_m$ are the $m$-th eigenstate and the energy level of the cluster Hamiltonian. A direct evaluation leads to $\lambda_c =1$
in any dimension, marking the quench of the magnetic order. The independence of the critical value on the dimensionality reflects the peculiar multi-spin interaction of the cluster couplings.

\section{Conclusions}
We studied the phase diagram and the geometric entanglement of the (spin $1/2$) cluster Hamiltonian in the presence of an additional Ising
interaction. We showed that Cluster states have symmetry protected topological order, which can be quenched to a long range order by varying the two spin interaction in a continuous way.
The one dimensional model we  studied in this paper is equivalent to a free fermion Hamiltonian, and therefore it can be solved exactly.  The Hamiltonian  enjoys a duality symmetry,  mapping the three point correlations into nearest neighbor interactions; as a consequence the entanglement in the  cluster state results to be dual to the entanglement encoded into the Ising gs state. The
system is self-dual at the critical point. We have shown how the geometric entanglement characterizes the two phases and the QPT. The exact solution in one-dimension together with the argument we produced above for higher dimensions provides, in our opinion, a convincing scenario for the transition from
Neel order  to a cluster state ordering of topological nature.
The critical properties of the transition could be investigated by  exploiting the scheme developed in \cite{kay} .

These results are also relevant for quantum information technology.  It would be valuable, for example, to exploit  the Ising-cluster state duality relation as a resource for computational power in quantum algorithms. Finally, we observe that the cluster Hamiltonian can be realized in a triangular  optical lattice of two atomic species \cite{Pachos04}. The three-spin interactions arises from a kind of density-dependent tunneling of the atomic species.  Our results can therefore be tested in optical lattices and in trapped ions, where this model can be experimentally
realized.

\section{Acknowledgements}
We acknowledge M.A. Martin-Delgado and A. Miyake for useful comments.
This work is supported by the National Research Foundation \& Ministry of Education, Singapore and by the European Community (IP-EUROSQIP). Research at Perimeter Institute for Theoretical
Physics is supported in part by the Government of Canada through NSERC and
by the Province of Ontario through MRI. Wonmin Son acknowledge Sogang University for their hospitality.



\begin{thebibliography}{99}
\bibitem{goldenfeldbook}
	N. Goldenfeld, {\it Lectures on phase transitions and the renormalization group},
	(Addison Wesley, New York, 1992).
\bibitem{sachdevbook}
	S. Sachdev, {\it Quantum Phase Transitions},  (Cambridge University Press, Cambridge, 2000).
\bibitem{Wenbook}
	X. Wen, {\it Quantum Field Theory of Many-body Systems: From the Origin of Sound to an Origin
	of Light and Electrons},  (Oxford Univ. Press, New York, 2004).
\bibitem{deconfined} T. Senthil, {\it et al},  Science {\bf 303}, 1490 (2004).
\bibitem{sachdev-review}
	S. Sachdev, Nat. Phys. {\bf 4}, 173 (2008).
\bibitem{deconfined-firstorder}
	F.-J. Jiang {\it et al. },  J. Stat. Mech.  P02009 (2008); A. B. Kuklov  {\it et al. },  Ann. Phys. (N.Y.) {\bf 321}, 1602 (2006);
	A. B. Kuklov  {\it et al. },  Phys. Rev. Lett. 101, 050405 (2008).
\bibitem{sachdev-report}
	S. Sachdev, ``Exotic phases and quantum phase transitions: model systems and experiments",
	arXiv:0901.4103.
\bibitem{fradkin}
	K. S. Raman {\it et al.}, ``Quantum Magnetism", B. Barbara et. al. editors, (Springer Netherlands 2008).
\bibitem{Briegel01} H. J. Briegel and R. Raussendorf, Phys. Rev. Lett. {\bf 86}, 910 (2001).
\bibitem{Doherty09} A. C. Doherty and S. D. Bartlett, Phys. Rev. Lett. {\bf 103}, 020506 (2009).
\bibitem{Skrovseth09} S. O. Skr{\o}vseth and S. D. Bartlett, Phys. Rev. A {\bf 80}, 022316 (2009).
\bibitem{Nielsen05} M.A. Nielsen, Rep. Math. Phys. 57, 147 (2006).
\bibitem{Damian07} D. Markham, A. Miyake and S. Virmani, New J. Phys. {\bf 9}, 194 (2007).
\bibitem{Vedral08} V. Vedral, Nature {\bf 453}, 1004 (2008).
\bibitem{cirac-localizable} M. Popp {\it et al.}, Phys. Rev. A {\bf 71}, 042306 (2005).
\bibitem{camposvenuti} L. Campos Venuti, M. Roncaglia, Phys. Rev. Lett. {\bf 94}, 207207 (2005).
\bibitem{Wen-protected} X. Chen, Z-C. Gu, X-G. Wen, arxiv:1004.3835.
\bibitem{boundary} X. Wen, Advances in Physics, 44, 405 (1995); I. Affleck {\it et al.}, Phys. Rev. Lett. {\bf 59}, 799 (1987); Y. Kitaev, Sov. Phys. Usp. {\bf 44}, 131 (2001).
\bibitem{kitaev-leshouches} A. Kitaev and C. Laumann, ``Topological phases and quantum computation", arXiv:0904.2771.
\bibitem{haldane} F. D. M. Haldane, Phys. Rev. Lett. {\bf 61}, 1029 (1988); {\it Quantum Magnetism}, U. Schollw\"ock, {\it et al.} Eds.	(Springer, Berlin, 2004).
\bibitem{AKLT} F. Pollmann, E. Berg, A. M. Turner, M. Oshikawa, arXiv:0909.4059.
\bibitem{AKLT-protected} S-P. Kou and X-G. Wen, Phys. Rev. B 80, 224406 (2009); A. Miyake, arxiv:1003.4662.
\bibitem{Savit80} R. Savit, Rev. Mod. Phys. {\bf 52}, 453 (1980); E. Fradkin and L. Susskind, Phys. Rev. D {\bf 17}, 2637 (1978).		
\bibitem{Lieb61} E. Lieb,T. Schultz and D. Mattis, Ann. Phys. {\bf 16}, 407 (1961);
\bibitem{Bartlett06} S. D. Bartlett and T. Rudolph, Phys. Rev. A {\bf 74}, 040302 (R) (2006).
\bibitem{Amico08} L. Amico {\it et al.}, Rev. Mod. Phys. {\bf 80}, 517 (2008).	
\bibitem{Wei03} T. Wei and P. M. Goldbart, Phys. Rev. A {\bf 68}, 042307 (2003); T. Wei, D. Das {\it et al.}, Phys. Rev. A {\bf 71}, 060305(R) (2005).
\bibitem{Pachos04} J. K. Pachos and M. B. Plenio, Phys. Rev. Lett. {\bf 93}, 056402 (2004).
\bibitem{VBS-cluster} F. Verstraete and  J.I. Cirac,  Phys. Rev. A {\bf 70}, 060302(R) (2004).
\bibitem{kay} A. Kay, Phys. Rev. Lett. {\bf 98}, 010501 (2007).
\bibitem{negele-orland} J. W. Negele and H. Orland, {\it Quantum Many-Particle Systems}, (Addison-Wesley 1988).
\bibitem{cya} L. Amico et al, in preparation.
\end{thebibliography}
\end{document}